\documentclass[iop]{emulateapj}
\usepackage[fleqn]{amsmath}
\usepackage{color}
\usepackage{hyperref}
\usepackage{url}
\hypersetup{
pdftitle={Identification of the white dwarf companion to millisecond pulsar J2317+1439},
pdfauthor={Shi Dai},
colorlinks=true,
linkcolor=blue,
citecolor=blue,
urlcolor=blue
}

\begin{document}

\title{Identification of the white dwarf companion to millisecond pulsar J2317$+$1439}

\author{S. Dai$^{1}$, M. C. Smith$^{2}$, S. Wang$^{3}$, S. Okamoto$^{2}$, R. X. Xu$^{4}$, Y. L. Yue$^{3}$ and J. F. Liu$^{3}$}
\affil{$^{1}$CSIRO Astronomy and Space Science, Australia Telescope National Facility, Box 76 Epping NSW 1710, Australia; shi.dai@csiro.au\\
$^{2}$Shanghai Astronomical Observatory, Chinese Academy of Sciences, Shanghai 200030, China\\
$^{3}$National Astronomical Observatories, Chinese Academy of Sciences, Beijing 100012, China\\
$^{4}$School of Physics and Kavli Institute for Astronomy and Astrophysics, Peking University, Beijing 100871, China}

\begin{abstract}
We report identification of the optical counterpart to the companion 
of the millisecond pulsar J2317+1439. At the timing position of the pulsar, 
we find an object with $g=22.96\pm0.05$, $r=22.86\pm0.04$ and $i=22.82\pm0.05$. 
The magnitudes and colors of the object are consistent with it being a white dwarf.
By comparing with white dwarf cooling models, we estimate that it 
has a mass of $0.39^{+0.13}_{-0.10}$\,M$_{\odot}$, an effective temperature of 
$8077^{+550}_{-470}$\,K and a cooling age of $10.9\pm0.3$\,Gyr. 
Combining our results with published constraints on the orbital parameters obtained 
through pulsar timing, we estimate the pulsar mass to be $3.4^{+1.4}_{-1.1}$\,M$_{\odot}$. 
Although the constraint on the pulsar mass is still weak, there
is a significant possibility that the pulsar could be more massive
than two solar mass. 
\end{abstract}


\section{Introduction} \label{sec:intro}

Millisecond pulsars (MSPs) are a special subgroup of radio pulsars, 
with shorter spin periods and much smaller spin-down rates compared 
to `normal' pulsars. 
Most MSPs have a low-mass white dwarf (WD) companion, and their fast
spins are believed to be a result of mass transfer from the
progenitor of the WD, known as recycling~\citep[e.g.,][]{tau11}.
Measuring the masses of MPSs and their companions allows us to study
these systems in detail, learning about their formation, evolution and
the accretion process. 
Mass measurements of pulsars also enable constraints to be
placed on the state of ultra-dense matter~\citep{dpr+10,afw13} and,
together with radio observations, can be used to test general relativity
~\citep[e.g.,][]{ksm06,s14}. 
Precise masses of MSPs and their companions can be determined through
high precision pulsar timing by measuring the Shapiro delay, but this
is possible only in exceptional cases. An alternative way to achieve
this relies on combined optical and radio timing
observations~\citep[e.g.,][]{vbk96}. For WD companions bright enough
for optical spectroscopy, a comparison of their spectrum with WD
atmosphere models can determine the effective temperature and surface
gravity. These can then be compared to WD evolutionary models to
obtain their masses. 
The mass ratio can be determined through pulsar timing and/or spectroscopy 
of the WD (using the amplitude of the radial-velocity curve), which can then be combined
with the WD mass to reveal the pulsar mass~\citep[e.g.,][]{vbj05}.

PSR J2317$+$1439 is a 3.4\,ms pulsar in a 2.46\,day orbit~\citep{cnt93}. 
The extremely low eccentricity of this binary system allows a tight 
test of the local Lorentz invariance of gravity~\citep{bcd96}.
Through long-term pulsar timing, the parallax of this pulsar has been 
measured to be $0.7\pm0.2$\,mas~\citep{mnf+16}. 
Shapiro delay effects caused by the companion have been observed
through high precision pulsar timing of the MSP, but these are weak
and produce relatively poor constraints on the masses of the companion
and the MSP~\citep{fpe+16}.

Previously, the companion to PSR J2317$+$1439 has not been reliably identified. 
\citet{mcp14} reported an association between the pulsar and a faint 
SDSS source J231709.23$+$143931.2, which has the following magnitudes:
$u>23.3$, $g=22.95\pm0.16$, $r=23.09\pm0.25$, $i>22.9$, $z>25.5$.
However, because this object is so faint, the SDSS photometry has large
uncertainties and hence it is difficult to ascertain the nature of the source.
In this paper, we report our optical identification of the 
companion to PSR J2317$+$1439 with the {\it Canada-France-Hawaii Telescope} (CFHT). 
We estimate the temperature, age and mass of the companion based on  
WD cooling models and constrain the possible mass of the MSP.
The identification of the companion opens up the prospect of optical
spectroscopy, leading to precise mass measurements for both the MSP and
WD. In turn this could lead to more stringent tests of gravity
theories and new constraints on the equation of state of pulsars.

Details of the observations and data analysis are given in 
Section \ref{sec:observations}. We estimate the mass of WD and pulsar
in Section \ref{sec:modelling}. A summary of our results and
discussions are given in Section \ref{sec:conclusion}.

\section{Observational Data}
\label{sec:observations}

\subsection{Observations and Data Reduction}

We used the MegaCam on CFHT to take $g$-, $r$- and $i$-band images of
a 1x1 square degree field containing PSR J2317$+$1439. This CFHT
program (12BS08; PI S. Dai) was applied through the Chinese Telescope
Access Program\footnote{\url{http://info.bao.ac.cn/tap/}}. The data
were taken from July 15 to 20 in 2012 for the three bands, with an
additional $g$-band observation in September 17 of that year. The
total exposure time was 1000, 2400 and 4300\,s for the $g$-, $r$- and
$i$-bands, respectively, with seeing between 0.{\arcsec}8 to 1.{\arcsec}0.
Each filter's observation was split into multiple exposures to avoid
saturation of bright stars, and dithered slightly between exposures to
span the gaps between chips and to correct for bad pixels.

\begin{figure*}
\begin{center}
\includegraphics[width=17cm]{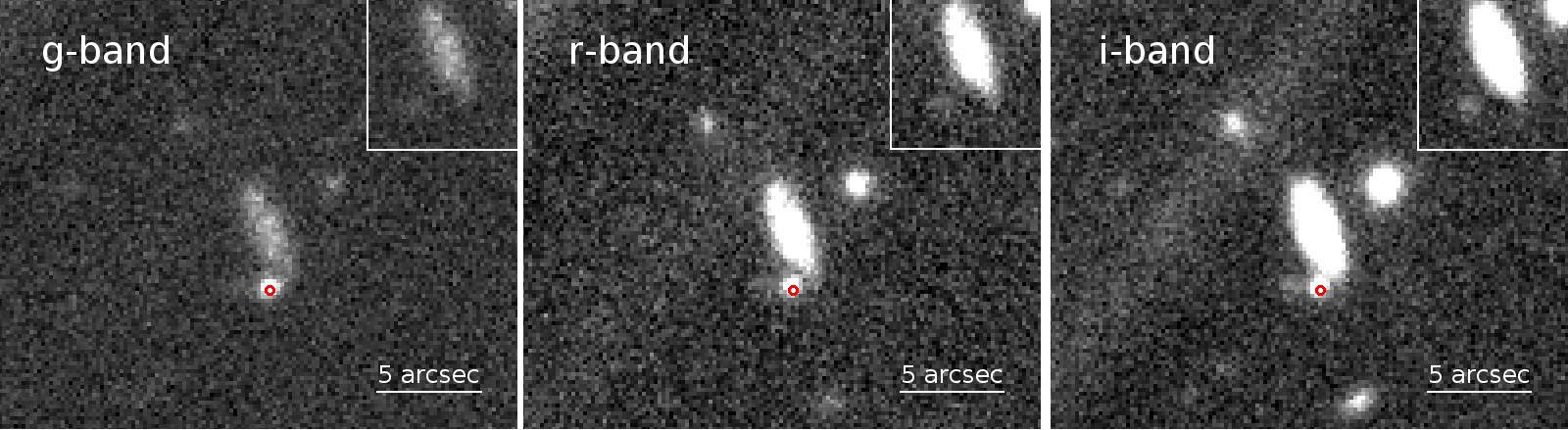}  
\caption{Cutouts showing our CFHT data around the location of PSR J2317$+$1439. 
The timing position of the MSP is marked with a red circle and the radius
corresponds to our astrometric uncertainty (0.{\arcsec}2). Cutouts of the PSF subtracted 
images are shown in the upper right corner for each band. The grey scale of 
each image shows the same luminosity range. Note that the faint diffuse diagonal 
bands in some of these images are a ghost from a nearby bright star.}
\label{images}
\end{center}
\end{figure*}

The data were pre-processed at CFHT with the Elixir pipeline\footnote{\url{http://www.cfht.hawaii.edu/Instruments/Imaging/MegaPrime/}} 
to correct for the instrumental signature across the whole mosaic.
The pre-processed data were then processed at
Terapix\footnote{\url{http://terapix.iap.fr/}}
with a pipeline that has been used for the CFHT Legacy
Survey\footnote{\url{http://www.cfht.hawaii.edu/Science/CFHTLS/}}.
The initial photometric calibrations were derived with \textit{Scamp}~\citep{b06} 
using the Ninth Sloan Digital Sky Survey (SDSS) Data Release (DR9).
Astrometric calibration was performed as a part of the
pipeline\footnote{\url{http://terapix.iap.fr/cplt/T0007/doc/T0007-doc.html}}
using the 2MASS catalogue.
The resulting astrometric uncertainties are 0.{\arcsec}23 in right
ascension and 0.{\arcsec}21 in declination, using 1515 bright
objects identified in both our images and 2MASS catalogue.
Once aligned astrometrically, exposures were rescaled and co-added by
\textit{Swarp}~\citep{bmr+02} using the \textit{Scamp} initial
photometric rescaling. Subsections of the co-added images containing
PSR J2317$+$1439 are shown in Fig.~\ref{images}.

\subsection{Photometry}
\label{sec:phot}

We performed point spread function (PSF) photometry of the candidate MSP
companion star, as well as of the field stars, using the co-added
images. This was done using the DAOPHOT II package~\citep{ste94}, which is
distributed as a part of the IRAF software environment.
We first used task \textit{daofind} to obtain a coordinate 
list of detected objects through the analysis of the co-addded images. Then we 
performed aperture photometry with task \textit{phot}. Task \textit{pstselect} 
was used to select 300 isolated, bright, unsaturated stars across 
the field, and task \textit{psf} was used to produce reliable PSF models for  
images of all three bands. We set \textit{varorder=2} to allow the PSF model 
to vary over the image. PSF-fitting photometry was then performed with task 
\textit{allstar} to obtain magnitudes and errors of objects in the list. 

\begin{figure}
\begin{center}
\includegraphics[width=3.5 in]{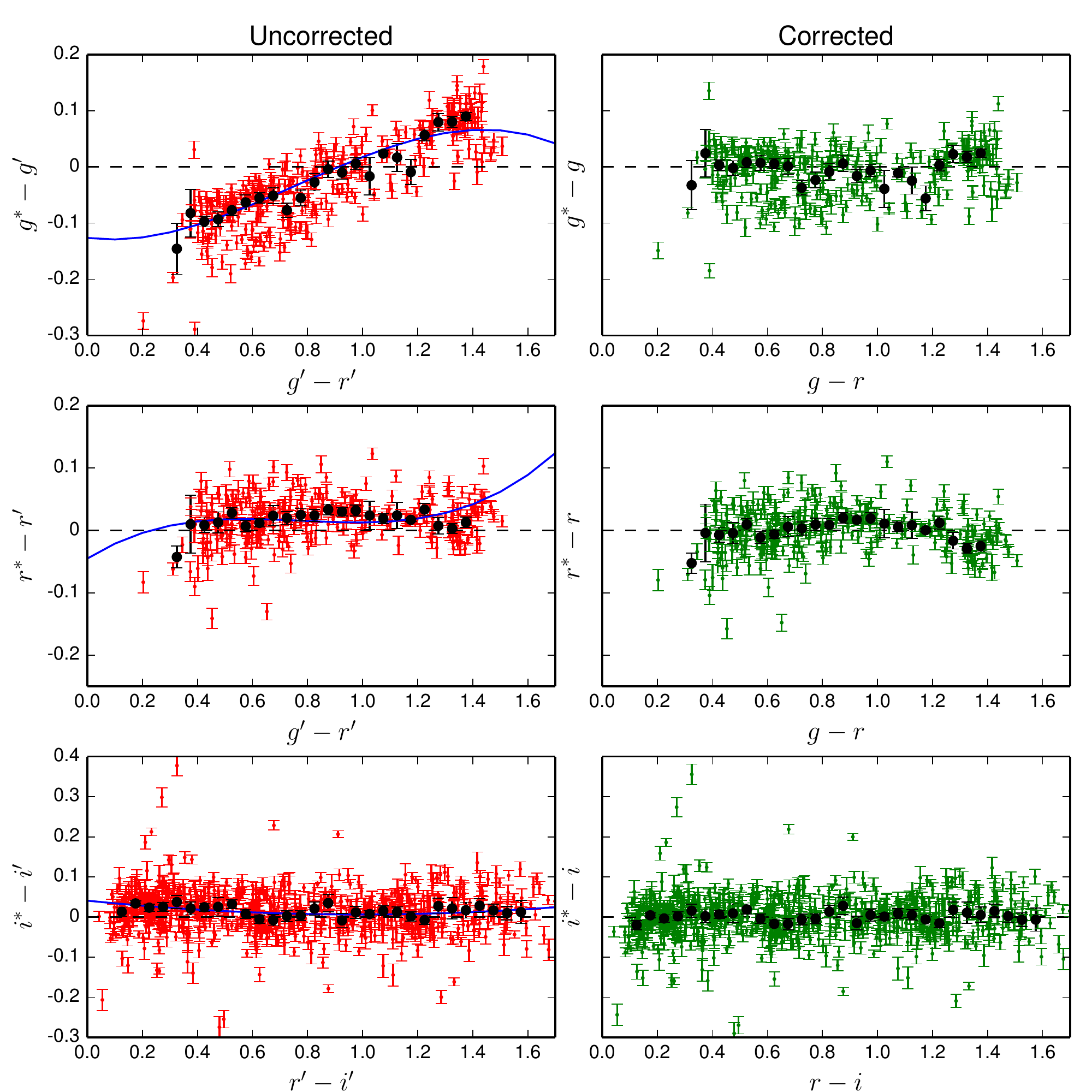}
\caption{Dependence of the $m_{\rm{sdss}}-m_{\rm{cfht}}$ residuals on 
CFHT color. Left panels show the uncorrected residuals and right panels show the 
corrected results. Black points represent the median of residuals and 
blue lines show the best-fit polynomials.}
\label{colordep}
\end{center}
\end{figure}

We recalibrated the photometry against SDSS DR9, fitting for the zero-points 
with 423, 580 and 708 isolated, unsaturated ($17<m_{\rm{sdss}}<20$),
point sources (sharpness parameter $|sh|<0.5$) selected in the
$g$-, $r$- and $i$-bands, respectively.
There was a clear dependence of the $m_{\rm{sdss}}-m_{\rm{cfht}}$ residuals on  
CFHT colors, most significantly in the $g$-band (for which the residual was
as much as 0.1 mag). To correct such color dependences, we used
transformations based on \citet{sji11}. For the $g$- and $r$-band
magnitudes we used equations (7) and (8) of 
\citet{sji11}\footnote{Note that the cubic terms are missing from
equations (7) and (8) of \citet{sji11}; these should be
$-0.15920082$ and $+0.16278071$ for (7) and (8), respectively
(B. Sesar, private communication).}, but kept the constant terms as a
free parameter. To determine the value of this term for each band, we
fitted the medians of $m_{\rm{sdss}}-m_{\rm{cfht}}$ for our
cross-matched stars. The best-fit values for the constant terms were
-0.127 and -0.045 for the $g$- and $r$-bands, respectively. We found
that the $i$-band residuals were not well-fit by the relation from
\citet{sji11}, and so we fit those ourselves using a quadratic
polynomial.  
The best-fit polynomials gave us the following transformations, where $g'r'i'$ 
and $gri$ correspond to uncalibrated and calibrated magnitudes, respectively:
\begin{equation}
g = g' - 0.127 - 0.062\times(g'-r') + 0.365\times(g'-r')^{2}
\end{equation}
\[
- 0.159\times(g'-r')^{3},
\]
\begin{equation}
r = r' - 0.045 + 0.275\times(g'-r') - 0.380\times(g'-r')^{2}
\end{equation}
\[
 + 0.163\times(g'-r')^{3},
\]
\begin{equation}
i = i' + 0.042 - 0.078\times(r'-i') + 0.041\times(r'-i')^{2}.
\end{equation}
In the left panels of Fig.~\ref{colordep}, we show the uncorrected $m_{\rm{sdss}}-m_{\rm{cfht}}$ 
residuals, where $m_{\rm{sdss}}=g^{*},r^{*},i^{*}$ are the SDSS magnitudes and $m_{\rm{cfht}}=g',r',i'$ 
are the calibrated CFHT magnitudes. Black points show the median of residuals and 
blue lines represent the best-fit polynomials. In the right panels, we show the color 
dependence corrected residuals and their medians, where $m_{\rm{cfht}}=g,r,i$ are the 
corrected CFHT magnitudes.

The scatter in the $m_{\rm{sdss}}-m_{\rm{cfht}}$ residuals is around
0.05-0.1 mag, which is larger than the internal errors and indicates
that there are systematic uncertainties remaining in our
photometry. To account for this, we first iteratively clipped 
$3\sigma$ outliers and calculated the standard deviations of
$m_{\rm{sdss}}-m_{\rm{cfht}}$ residuals in each band ($\sigma_{\rm{res}}$).
This dispersion is a combination of the systematic uncertainties
($\sigma_{\rm{sys}}$) together with the internal uncertainties from
our CFHT data ($\sigma_{\rm{cfht}}$) and from SDSS
($\sigma_{\rm{sdss}}$), i.e. the systematic uncertainty can be approximated by
\begin{equation}
  \sigma_{\rm{sys}}=\sqrt{\sigma_{\rm{res}}^{2}-\sigma_{\rm{cfht}}^{2}-\sigma_{\rm{sdss}}^{2}}.
\end{equation}
These systematic uncertainties are listed in Table \ref{pars} for each
band.

\begin{table*}
\begin{center}
\caption{Parameters of PSR J2317$+$1439 and photometric results of the
companion. The timing, astrometric and orbital parameters are from \citet{dcl+16}, \citet{mnf+16} and \citet{fpe+16}. 
The extinction ($A_{\lambda}$) is estimated using models by \citet{gsf+15} and coefficients from \citet{sf11}.}
\label{pars}
\begin{tabular}{lccc}
\hline
\hline
\multicolumn{4}{c}{Timing parameters~\citep{dcl+16}}   \\
\hline
$P_{\rm{s}}$ (ms)                                  &   \multicolumn{3}{c}{3.44525112564488(18) }                          \\
$\dot{P}_{\rm{s}}$ (10$^{-20}$\,s\,s$^{-1}$)       &   \multicolumn{3}{c}{0.2433(3)}                                      \\
$\tau$ (10$^{9}$\,yr)                              &   \multicolumn{3}{c}{15.6}                                           \\
\hline
\multicolumn{4}{c}{Astrometric parameters~\citep{mnf+16}}   \\
\hline
$\alpha_{\rm{J2000}}$                              & \multicolumn{3}{c}{23$^{\rm{h}}$17$^{\rm{m}}$09$^{\rm{s}}$.236644(9)} \\
$\delta_{\rm{J2000}}$                              & \multicolumn{3}{c}{$+$14$^{\circ}$39$^{'}$31.{\arcsec}2557(2)}          \\
$\mu_{\alpha}$                                     & \multicolumn{3}{c}{-1.39(3)\,mas\,y$^{-1}$}                           \\
$\mu_{\delta}$                                     & \multicolumn{3}{c}{3.55(6)\,mas\,y$^{-1}$}                           \\
Parallax                                           & \multicolumn{3}{c}{0.7(2)\,mas}                                       \\
\hline
\multicolumn{4}{c}{Orbital parameters~\citep{fpe+16}}   \\
\hline
$P_{\rm{b}}$                                       & \multicolumn{3}{c}{2.45933146519(2) (days)}                            \\
$x$                                                & \multicolumn{3}{c}{2.313943(4) (lt-s)}                                 \\
$i$                                                & \multicolumn{3}{c}{47$^{+10}_{-7}$ (deg) }                             \\
$e$                                                & \multicolumn{3}{c}{$5.7(16)\times10^{-7}$ }                            \\
\hline
\multicolumn{4}{c}{Photometric results}   \\
\hline
                   & g-band             &     r-band          &      i-band      \\
Magnitudes         & 22.96 $\pm$ 0.02 & 22.86 $\pm$ 0.03  & 22.82 $\pm$ 0.03 \\
$\sigma_{\rm{sys}}$          & 0.04     & 0.03                & 0.04              \\
$A_{\lambda}$                & 0.185    & 0.128               & 0.095              \\
\hline
\end{tabular}
\end{center}
\end{table*}

\subsection{Identification of the optical companion to PSR J2317$+$1439}

We identified an optical object at the timing position of the MSP in 
all three bands. The optical position is
$\alpha_{\rm{J2000}}=23^{\rm{h}}17^{\rm{m}}09^{\rm{s}}.24$ 
and $\delta_{\rm{J2000}}=14^{\circ}39\arcmin31.\arcsec46$, with an
uncertainty of around 0.\arcsec2 in each coordinate coming from the
astrometric calibration.
The timing and astrometric parameters of the MSP are listed in
Table~\ref{pars}~\citep{dcl+16, mnf+16} and the offset with our
detection is around 0.\arcsec24, i.e. consistent with the uncertainty
in the astrometric calibration. 
The reference epoch of astrometric parameters is $\rm{MJD}=55000$, and 
the offsets introduced by pulsar proper motions at epochs of our 
optical observations are $\Delta\alpha\approx-4.2$\,mas and $\Delta\delta\approx10.7$\,mas, 
which are negligible compared with astrometric uncertainties of the optical 
position.
The astrometry of our detection also agrees with that of the SDSS 
object identified by \citet{mcp14}.
For objects with $g<24$\,mag, CFHT images have an average stellar density of six
stars per square arcminute, which translates to only a 0.02 per cent
probability of a chance coincidence within an error circle with a 
radius of 0.\arcsec2.
In Fig.~\ref{images}, we show cutouts of the CFHT images, with the 
timing position of the MSP marked as red circle with a radius of 0.{\arcsec}2. 

As can be seen from these images there is a background galaxy lying
close to our optical object. However, this should not affect our
photometry because the object is clearly resolved in all bands and
we used PSF photometry.

The distance to the pulsar is estimated to be $D_{\rm{psr}}=1.3^{+0.4}_{-0.3}$\,kpc 
based on the parallax measurement. We used the Bayesian approach described in Eq. 22 
of \citet{ivc16}. As summarised in the lower part of Table 1 from \citet{ivc16}, the 
priors assume a pulsar density distribution\footnote{Note that Table 1 of \citet{ivc16} 
does not use the same scale-height as the reported reference \citep{lfl+06}. We use 
the value directly from \citet{lfl+06}, which is $h=0.33$\,kpc.} from \citet{lfl+06}, 
luminosity function from \citet{fk06}, and we take the flux to be $4\pm1$\,mJy 
at 1.4\,GHz~\citep{kxl+98}.

The magnitudes of the optical object are listed in Table~\ref{pars}.
Note that these magnitudes are after applying the color-dependent
correction described in Section \ref{sec:phot}, but before applying
any extinction correction. Using models by~\citet{gsf+15}, we obtained a 
reddening $E(B-V)$ of $0.056\pm0.03$\,mag for a distance of 1.3\,kpc towards 
PSR J2317$+$1439.
Combined with the $R_{\rm{V}}=3.1$ extinction law and coefficients for
SDSS filters from~\citet{sf11}, the $g$-, $r$- and $i$-band extinctions 
are estimated and given in Table \ref{pars}.

\begin{figure*}
\begin{center}
\includegraphics[width=6.5 in]{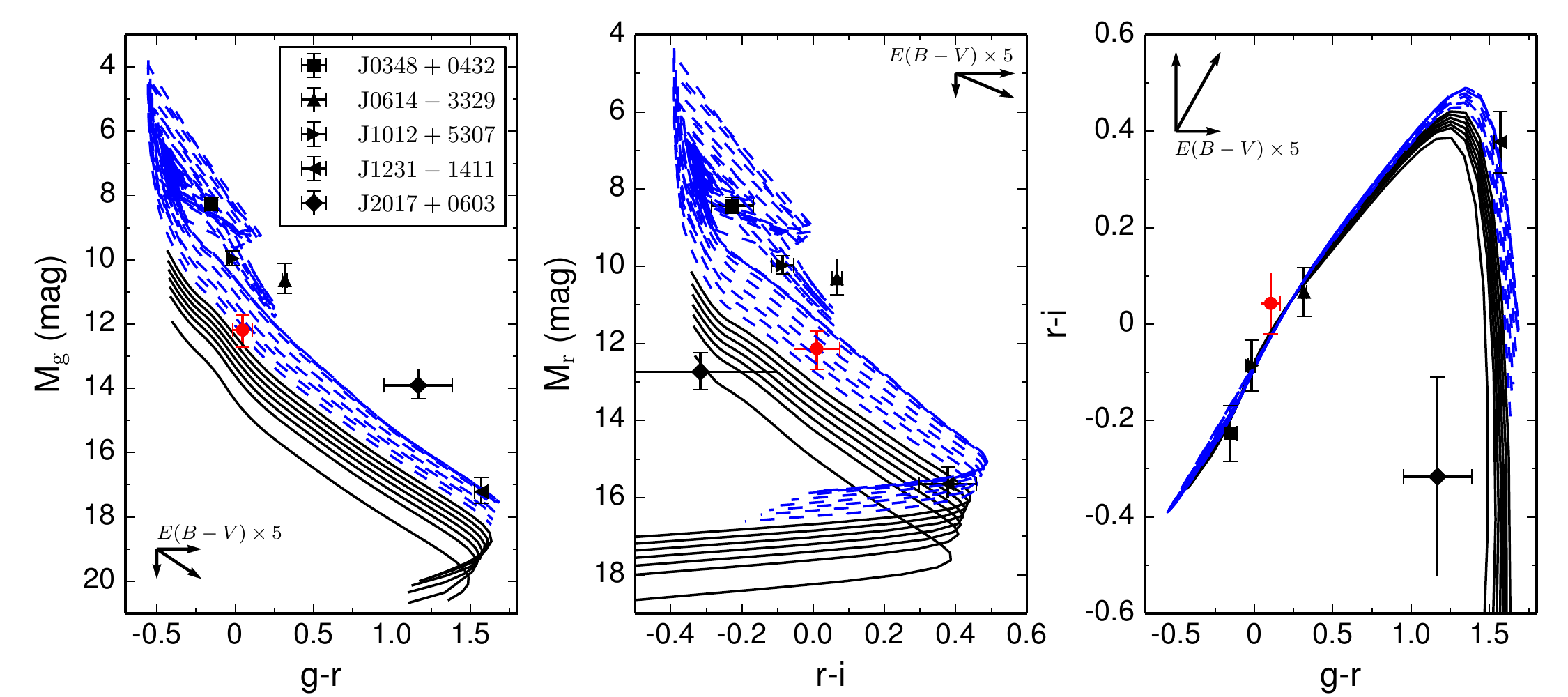}
\caption{Color-magnitude diagrams and color-color diagram. In the
  magnitude-color diagrams, absolute magnitudes (estimated using
  $D_{\rm{psr}}=1.3^{+0.4}_{-0.3}$\,kpc) are shown as red points
  with error bars. Solid black lines show CO-core WD models, with 
  masses varying linearly from 0.5 to 1.2\,$\rm M_\odot$. Dashed 
  blue lines show ELM WD models from \citet{amc13}, with 
  masses varying linearly from 0.1554 to 0.4352\,$\rm M_\odot$. 
  Magnitudes and colors of the companions to PSRs J0348$+$0432, 
  J0614$-$3329, J1012$+$5307, J1231$-$1411 and J2017$+$0603 are 
  shown as black points with error bars. For PSR J2317$+$1439, the 
  estimated reddening is $E(B-V)=0.056\pm0.03$\,mag, and we have 
  included the reddening vector on each panel (scaled up by a factor 
  of five for clarity).}
\label{he_gr}
\end{center}
\end{figure*}

The dereddened color-magnitude and color-color diagrams are presented
in Fig.~\ref{he_gr}. The absolute magnitudes are estimated using
$D_{\rm{psr}}=1.3^{+0.4}_{-0.3}$\,kpc and the corresponding
uncertainties are dominated by the distance uncertainties. 
Most WD companions to MSPs are known to be low-mass helium-core WDs 
with masses below 0.2--0.3\,M$_{\odot}$~\citep[e.g.,][]{vbj05} and 
are called extremely low-mass (ELM) WDs. 
In Fig.~\ref{he_gr}, we compare our magnitudes and colors with theoretical 
evolutionary tracks for ELM WD models, covering WD masses from $\sim0.16$ to 0.44\,M$_{\odot}$.
The ELM WD cooling models come from \citet{amc13}\footnote{\url{http://evolgroup.fcaglp.unlp.edu.ar/TRACKS/tracks_heliumcore.html}},
where theoretical luminosities and temperatures have been transformed  
into absolute magnitudes by applying bolometric corrections for pure hydrogen 
model atmospheres~\citep[provided by P. Bergeron, see][]{hb06,bwd+11}.
MSPs with more massive WD companions (e.g., PSR J1614$-$2230, \citealt{dpr+10}) 
have also been found and are proposed to evolve from intermediate-mass X-ray binaries~\citep[e.g.,][]{tlk11}.
Therefore, we also consider evolutionary tracks for carbon-oxygen (CO) core 
WDs with pure hydrogen model atmospheres, covering WD masses from 0.5 to 1.2\,M$_{\odot}$. 
These models are from \citet{hb06}, \citet{ks06}, \citet{tbg11} and \citet{bwd+11}\footnote{See \url{http://www.astro.umontreal.ca/~bergeron/CoolingModels/} for more details about cooling models and color calculations.}.

The magnitudes and colors of our source are in good agreement with
the ELM models, but lie at the low mass side of the CO-core WD models. 
Although the colors of the object are also consistent with other blue 
stars, such as blue-horizontal branch or blue-straggler stars, the 
magnitudes would imply a distance of many kpc in which case it could 
not be associated to the pulsar. 
For comparison, in Fig.~\ref{he_gr} we also presented magnitudes and colors of the companions to 
PSRs J0348$+$0432~\citep{afw13}, J0614$-$3329~\citep{bac+16}, J1012$+$5307~\citep{nll+95}, 
J1231$-$1411~\citep{bac+16} and J2017$+$0603~\citep{bac+16}\footnote{Apparent 
magnitudes of the companions to PSRs J0348$+$0432 and J1012$+$5307 were obtained 
from the Sloan Digital Sky Survey~\citep{yaa+00} webpage (\url{http://skyserver.sdss.org/dr13/}).
For PSRs J0614$-$3329, J1231$-$1411 and J2017$+$0603, the distances are not well 
constrained and we used distances estimated from dispersion measures~\citep{bac+16} 
and assumed 20\% uncertainties~\citep{cl02}.}. Extinctions have been corrected following 
the same procedure as for PSR J2317$+$1439.

\section{Estimating the mass of the companion and pulsar}
\label{sec:modelling}

Since we have both the colors and distance to the companion, we can
use models to constrain the mass, temperature and age of the WD. 
We have done this by constructing a single composite model which uses the ELM tracks 
for the mass range 0.1554 to 0.4352\,$\rm M_\odot$ and CO-core tracks for 
the mass range 0.5 to 1.2\,$\rm M_\odot$.	
We interpolated these models in the mass--temperature plane using 
natural neighbour interpolation with the IDL command `griddata'.

Assuming Gaussian errors on the photometry, the likelihood of any
given model point is described by the following equation,
\begin{equation}
  \mathcal{L}
  =\prod_{f=g,r,i} \frac{1}{\sqrt{2\pi\delta_f^2}}
  \,{\rm exp}\left(\frac{-(m_f-m_f^{\rm model})^2}{2\delta_f^2}\right),
\end{equation}
where $m_f$ and $\delta_f$ are the apparent magnitude and error for
our observed bands $f=g,r,i$ and the model is a function of the unknown
parameters (in our case effective temperature, WD mass, and
distance). We calculated the likelihood using this equation for each
point in our 2D interpolated plane, taking a $4000\times4000$ grid linearly
spaced in the temperature range 6,000 to 10,000\,${\rm K}$ and in the 
mass range 0.1554 to 1.2\,${\rm M_\odot}$.

\begin{figure*}
\begin{center}
\includegraphics[width=4 in]{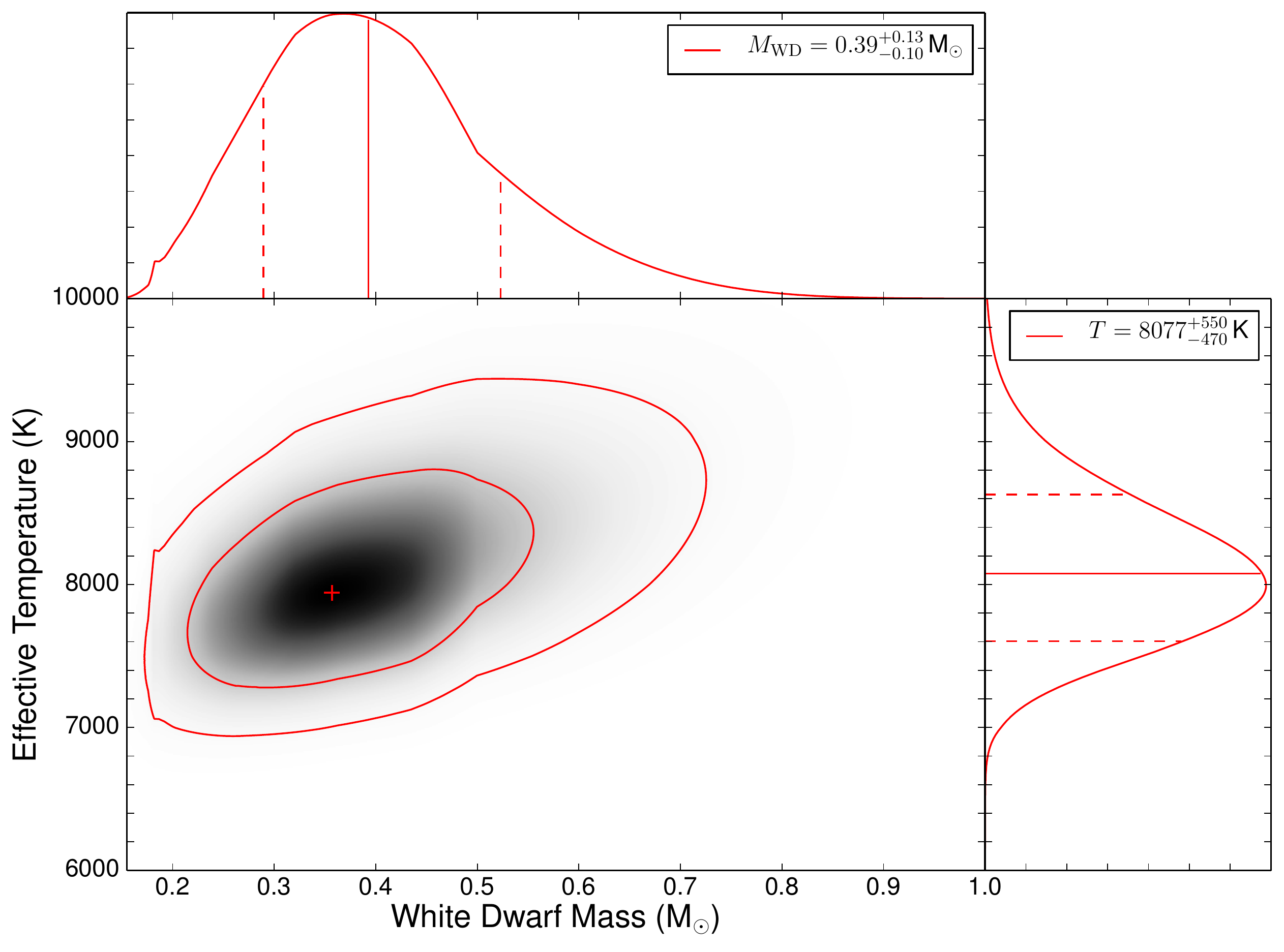}
\caption{Constraints on the WD mass and effective temperature from the
CFHT photometry using the composite ELM and the CO-core WD models. The contours 
correspond to 1- and 2-$\sigma$ confidence intervals and the peak is denoted by 
a cross. The marginalized 1D likelihoods are presented in
the top and side panels, with the solid and dashed lines showing the
median and 1$\sigma$ confidence intervals, respectively.}
\label{fig:wd_mass}
\end{center}
\end{figure*}

As outlined in Section 2.3, we have used Eq. 22 of \citet{ivc16} to 
estimate the pulsar distance; we use the resulting probability distribution 
function as a prior in Eq. 2.
We correct our magnitudes for extinction, as discussed in Section 2.3, 
and incorporate the 0.03\,mag uncertainty on the reddening in our modeling. 
We used uniform priors on both effective temperature and WD mass. The
resulting constraints on the effective temperature and WD mass are shown 
in Fig.~\ref{fig:wd_mass}. 
We obtained a WD mass of $0.39^{+0.13}_{-0.10}$\,M$_{\odot}$, 
an effective temperature of $8077^{+550}_{-470}$\,K and a cooling age 
of $10.9\pm0.3$\,Gyr, where we have quoted the median of the probability 
distribution and the $1\sigma$ error. 

\begin{figure}
\begin{center}
\includegraphics[width=3.3 in]{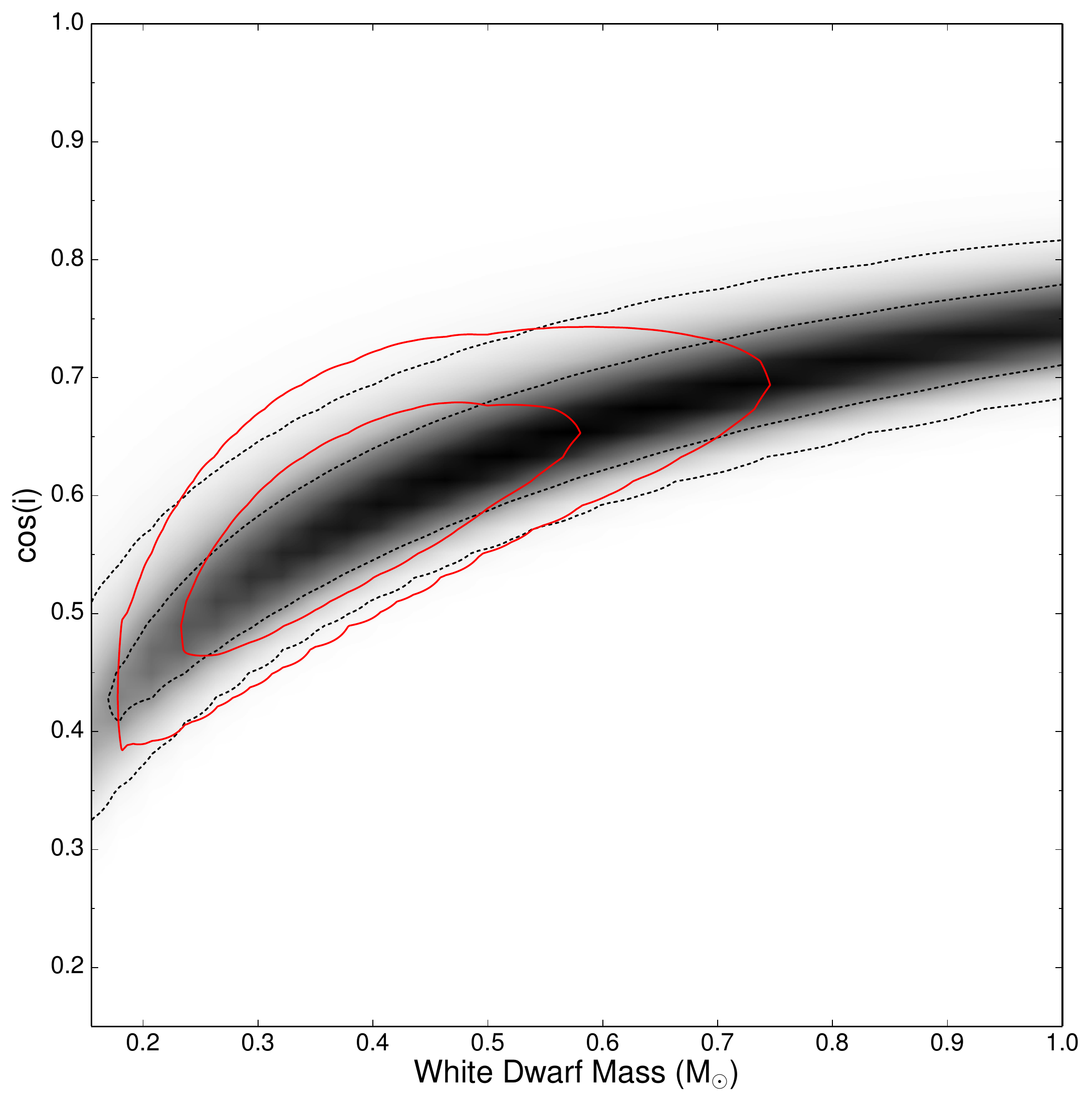}
\caption{Constraints on the WD mass and inclination angle of the binary system. 
The grey-scale and dashed contours correspond to the constraints derived from PSR 
timing~\citep{fpe+16}, while the solid contours show the constraints after applying 
a prior on the WD mass derived from our CFHT photometry and WD models.}
\label{fig:ns_mass}
\end{center}
\end{figure}

\begin{figure}
\begin{center}
\includegraphics[width=3.3 in]{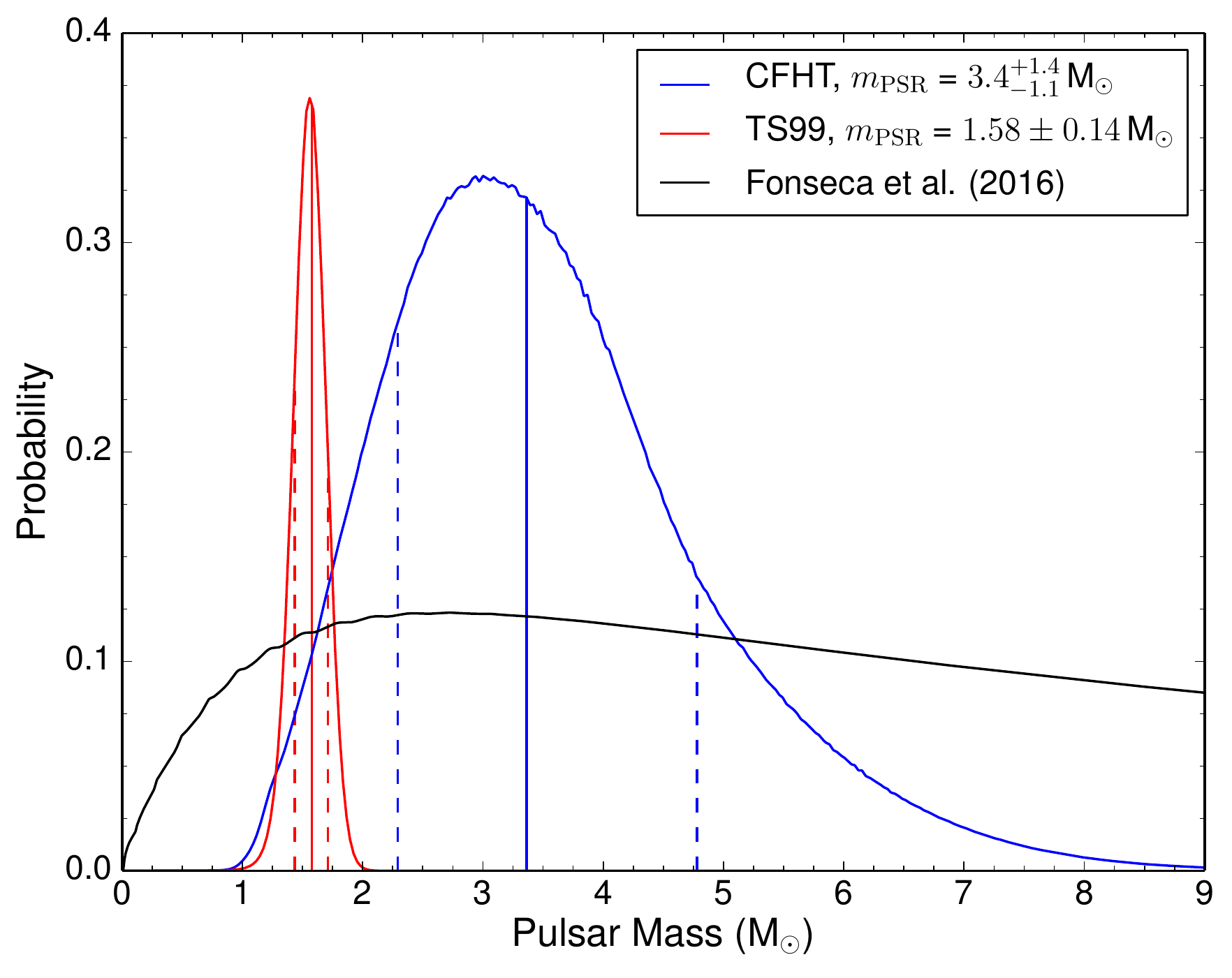}
\caption{Constraints on the mass for PSR J2317$+$1439. 
Each curve is normalized so that the area underneath is unity, 
except the red curve which has been scaled down by a factor of six.
The vertical solid and dashed lines denote the median and 1$\sigma$
confidence intervals, respectively.}
\label{fig:ns_like}
\end{center}
\end{figure}

Our constraints on the WD mass can be used to further constrain the
pulsar mass through the equation,
\begin{equation}
\frac{(m_{\rm{WD}}\sin i)^3}{(m_{\rm{PSR}}+m_{\rm{WD}})^2}=\frac{4\pi^{2}}{G}\frac{x^3}{P_{\rm{b}}^2},
\label{eq:massFunc}
\end{equation}
where $i$ is the inclination angle, $x$ is the projected semi-major
axis and $P_{\rm{b}}$ is the orbital period. The most up-to-date
estimates for the orbital parameters, which have been presented in
Table \ref{pars}, come from pulsar timing \citep{fpe+16}. For PSR J2317$+$1439 
the timing analysis leads to only weak constraints on the WD mass and, 
consequently, the pulsar mass. 
In Fig.~\ref{fig:ns_mass}, we show how the
timing confidence intervals (grey-scale and dashed contours) contract if we apply a
prior based on our photometric constraints on the WD mass (solid
contours). We can use these new constraints on the inclination and WD
mass to estimate the NS mass through Eq.~\ref{eq:massFunc}. 
The NS mass is now better constrained, with a 1$\sigma$
confidence interval of $3.4^{+1.4}_{-1.1}\,{\rm M_\odot}$ (see Fig. \ref{fig:ns_like}). 
Although this is still not a very tight constraint, it is indicative 
that the pulsar may be massive, with probabilities of only 9\% that 
the mass is below 2\,$\rm M_\odot$.

Previous studies have argued that the system of PSR J2317$+$1439 
has evolved from a low-mass binary, and has a helium-core WD
companion~\citep{vbj05}. The relation of WD mass to orbital period 
for systems evolved from low-mass binaries has 
been studied by a number of authors~\citep[e.g.,][]{ts99,lrp+11,imt+16}.
For orbital periods larger than 2\,days, previous studies gave 
very similar relations, which have been shown to agree well
with MSP binary systems with low-mass helium-core WD companions (see,
for example, figure 8 of \citealt{fpe+16}). 
For the 2.3\,day orbital period of PSR J2317$+$1439, assuming a helium-core 
WD companion, the \citet{ts99} models predict a WD mass of 0.21 to 0.23\,${\rm M_\odot}$,
where the spread comes from the uncertainty in the chemical abundance
of the WD. If we apply a Gaussian prior to the WD mass, with mean 0.22
and standard deviation 0.01 ${\rm M_\odot}$, the resulting pulsar
mass is $1.58\pm0.14\,{\rm M_\odot}$. 
The WD mass predicted by \citet{ts99} is inconsistent at 1$\sigma$ with our 
result. However, the current constraint on the pulsar 
parallax is not particularly tight and this is important because the WD mass 
is degenerate with its absolute magnitudes. To obtain a WD mass of 0.22\,${\rm M_\odot}$
the distance would need to be 1.94\,kpc, although this is outside the 1$\sigma$
constraint obtained in Section 2.3, a more precise measurement of the 
parallax would clearly reduce the uncertainty.

\section{Conclusion and discussion}
\label{sec:conclusion}

We have reported the optical identification of the companion to PSR
J2317$+$1439. The timing position of the pulsar agrees with the
optical position of the detection and the photometry agrees with WD
cooling models.
This identification opens up the possibility of precisely measuring
the WD temperature and surface gravity through optical spectroscopy,
although the faint nature of the star means that this will require
large optical telescopes. Combined with high precision pulsar timing,
this would lead to a precise mass measurement for the MSP.

By fitting the photometry with WD models, we have estimated the
mass of the WD to be $0.39^{+0.13}_{-0.10}$\,M$_{\odot}$ and the 
effective temperature to be $8077^{+550}_{-470}$\,K. 
The WD models predict a cooling age of $10.9\pm0.3$\,Gyr, 
which is close to the characteristic age of the pulsar of 15.6\,Gyr.
These estimates depend on the distance to the system, which can be
obtained from the trigonometric parallax measurement.
Since the parallax is not very well constrained ($0.7\pm0.2$ mas), the
Lutz-Kelker bias needs to be corrected for \citep[e.g.,][]{vwc+12} 
and we have incorporated the correction into our estimates following 
the Bayesian approach described in \citet{ivc16}. 

It has been suggested that this system has evolved from a low-mass
binary, and the companion is likely to be a helium-core
WD~\citep{vbj05}. Although our results agree with such a scenario, 
the WD mass of $0.39^{+0.13}_{-0.10}$\,M$_{\odot}$ is marginally inconsistent with 
predictions based on the relation of WD mass to orbital period.
For the 2.46\,days orbital period, models from \citet{ts99}
predict a WD mass of 0.21 to 0.23\,M$_{\odot}$, which is just outside the
1$\sigma$ confidence interval obtained from fitting our photometry
with WD models.
Therefore, the nature of the progenitor binary and how it evolved
during the mass-exchanging X-ray phase are still unclear. 

Combining our WD mass estimate with constraints on the orbital
parameters of this system derived from pulsar timing~\citep{fpe+16},
we have estimated the pulsar mass to be $3.4^{+1.4}_{-1.1}$\,M$_{\odot}$.
This is consistent with the mass measured by \citet{fpe+16}, but
with much smaller uncertainties. 
Although tentative, our results indicate that PSR J2317+1439 may 
be an extremely massive neutron star ($>$2.04\,M$_{\odot}$ at 90\% confidence).
If confirmed, this could challenge our understanding of the state 
of dense matter and structure of neutron stars~\citep[e.g.,][]{xg16}.
Long-term high precision timing of PSR J2317$+$1439 could in principle 
better measure the Shapiro delay and then the mass of both WD and pulsar, 
but this is limited by the timing precision we can achieve for this pulsar. 
However, further observations could also lead to an improved parallax measurement and
this would improve our WD mass estimate. For example, if the parallax
error was reduced by a factor of two to 0.1\,mas, then the
corresponding pulsar mass uncertainty would be reduced by around 25\%.
An alternative way is to obtain an optical spectrum of the WD, as
discussed previously. If one could measure the surface gravity of the
WD, this would dramatically reduce the allowed range of parameter
space and provide much tighter constraints on the pulsar mass.

\acknowledgments

The authors wish to thank E. Fonseca for providing his likelihood
distributions from pulsar timing, P. Bergeron for providing bolometric 
corrections and S. Justham for helpful comments.
This research uses data obtained through the Telescope Access Program
(TAP), which has been funded by the National Astronomical
Observatories of China, the Chinese Academy of Sciences (the Strategic
Priority Research Program ``The Emergence of Cosmological Structures"
Grant No. XDB09000000), and the Special Fund for Astronomy from the
Ministry of Finance.
M.C.S. acknowledges financial support from the CAS One Hundred Talent Fund, 
the National Key Basic Research Program of China 2014CB845700, and from NSFC 
grants 11173002 and 11333003.
R.X.X. acknowledges support from NSFC grants 11673002 and U1531243.
This work is based on data products produced at the TERAPIX data
center located at the Institut d'Astrophysique de Paris.
We thank all the people that have made this AASTeX what it is today.  This
includes but not limited to Bob Hanisch, Chris Biemesderfer, Lee Brotzman,
Pierre Landau, Arthur Ogawa, Maxim Markevitch, Alexey Vikhlinin and Amy
Hendrickson.




\end{document}